\documentclass[conference]{IEEEtran}

\usepackage[pdftex]{graphicx}
\usepackage{curves,epic,latexsym,amsmath,amssymb,array,cite,ifthen}
\usepackage{epsfig}

\newcommand{\eqdef}{\stackrel{\scriptscriptstyle\bigtriangleup}{=} }

\newcommand{\wH}{\mathrm{w_H}}

\newcommand{\dminH}{\mathrm{d_{minH}}}
\newcommand{\wD}{\mathrm{w_D}}
\newcommand{\dD}{\mathrm{d_D}}

\newcommand{\dminD}{\mathrm{d_{minD}}}

\newcounter{proglinecounter}
\newenvironment{pseudocode}%
    {\setcounter{proglinecounter}{0}%
     \begin{tabbing}12345\=12345\=123\=123\=123\=123\=123\=123\=123\=123\=123\= \kill}%
    {\end{tabbing}}
\newcommand{\npcl}[1][]
    {\>\refstepcounter{proglinecounter}\arabic{proglinecounter}%
     \ifthenelse{\equal{#1}{}}{}{\label{#1}}\' \>}
\newcommand{\pkw}[1]{\textbf{#1}}    

\newcounter{examplecntr}
{\begin{trivlist}\item[]\refstepcounter{examplecntr}%
 {\bfseries Example~\theexamplecntr%
  \ifthenelse{\equal{#1}{}}{}{ (#1)}.
}}%
{\hfill$\Box$\end{trivlist}}

\newcounter{definitioncntr}
\newenvironment{definition}%
{\begin{trivlist}\item[]\refstepcounter{definitioncntr}%
{\bfseries Definition~\thedefinitioncntr.}}%
{\hfill$\Box$\end{trivlist}}

\newcounter{theoremcntr}
\newenvironment{theorem}[1][]%
{\begin{trivlist}\item[]\refstepcounter{theoremcntr}%
{\bfseries Theorem~\thetheoremcntr%
  \ifthenelse{\equal{#1}{}}{}{ (#1)}.
}}%
{\hfill$\Box$\end{trivlist}}

\newcommand{\eproofnegspace}{\\[-1.5\baselineskip]\rule{0em}{0ex}}

\newcounter{propositioncntr}
{\begin{trivlist}\item[]\refstepcounter{propositioncntr}%
{\bfseries Proposition~\thepropositioncntr%
  \ifthenelse{\equal{#1}{}}{}{ (#1)}.
}}%
{\hfill$\Box$\end{trivlist}}

\begin{document}

\markboth{December 29, 2011}%
{Shell \MakeLowercase{\textit{et al.}}: Bare Demo of IEEEtran.cls
for Journals}

\title{On the Joint Error-and-Erasure Decoding for Irreducible Polynomial Remainder Codes}

\author{\IEEEauthorblockN{Jiun-Hung~Yu }
\IEEEauthorblockA{Department of Information Technology and
Electrical Engineering\\
ETH Zurich, Switzerland\\
Email: yu@isi.ee.ethz.ch}}

\maketitle

\begin{abstract}

A general class of polynomial remainder codes is considered. Such
codes are very flexible in rate and length and include
Reed-Solomon codes as a special case.

As an extension of previous work, two joint error-and-erasure
decoding approaches are proposed. In particular, both the decoding
approaches by means of a fixed transform are treated in a way
compatible with the error-only decoding. In the end, a collection
of gcd-based decoding algorithm is obtained, some of which appear
to be new even when specialized to Reed-Solomon codes.
\end{abstract}

\section{Introduction}
\label{Introduction}

Polynomial remainder codes, constructed by means of the Chinese
remainder theorem, were proposed by Stone \cite{Stone}, who also
pointed out that these codes include Reed-Solomon codes
\cite{Reed} as a special case. Variations of Stone's construction
were studied in \cite{Bossen,Mandelbaum,Mandelbaum2}, but no
efficient decoding algorithm for random error was presented in
these papers. There is also a connection between Goppa
\cite{Goppa} codes and polynomial remainder codes, as noted in
\cite{YuLoeliger2}.

In 1988, Shiozaki \cite{Shiozaki} proposed an efficient error-only
decoding algorithm for Stone's codes constructed by irreducible
moduli \cite{YuLoeliger2}. However, the algorithm is restricted to
codes with a fixed symbol size, i.e., fixed-degree moduli. This
restriction was overcome by the decoding algorithms in
\cite{YuLoeliger, YuLoeliger2}, which explicitly work for codes
with variable symbol sizes, i.e., variable-degree moduli. Note
that by admitting moduli of different degrees, a Reed-Solomon code
can be easily lengthened by adding some higher-degree symbols
without increasing the size of the underlying field
\cite{YuLoeliger, YuLoeliger2}. Note also that Shiozaki's
algorithm, when applied to Reed-Solomon codes, is the same as
Gao's \cite{Gao} algorithm as pointed out in
\cite{YuLoeliger2,LinChenTruong,Fedorenko2}.

In presence of both error and erasures, the error-only decoding
algorithm, as observed by Shiozaki \cite{Shiozaki}, can be applied
to shortened polynomial remainder codes interpolated by ignoring
the erased symbols. Such an interpolation, however, involves a lot
of re-computation of the interpolating basis and thus greatly
increases the decoding complexity. When applied to Reed-Solomon
codes, the same problem exists, as pointed out in
\cite{LinChenTruong}, but for Reed-Solomon codes, the problem can
be bypassed by the decoding algorithm of \cite{LinChenTruong}.

In this paper, we consider the extension of the error-only
decoding algorithms of \cite{YuLoeliger,YuLoeliger2} to joint
error-and-erasure decoding of irreducible polynomial remainder
codes. Two fixed-transform approaches are proposed for decoding
such codes. When applied to Reed-Solomon codes, the first approach
is essentially identical to the one in \cite{LinChenTruong}, but
the second approach appears to be new. For each approach, the
decoding algorithm consists of two steps: in the first step, a
polynomial which factorizes the error locator polynomial is
computed by means of a gcd algorithm; in the second step, the
message is recovered, for which we also propose two different
methods.

The paper is organized as follows. In Section~\ref{section:CRT},
we recall the Chinese remainder theorem and the definition of
irreducible polynomial remainder codes. In
Section~\ref{section:ErrorAndErasure}, we address the problem of
joint error-and-erasure decoding and propose two fixed-transform
decoding approaches. In Sections~\ref{section:exgcdI}
and~\ref{section:exgcdII}, we derive gcd-based decoding algorithms
for the respective approaches. A collection of these algorithms is
summarized in Section~\ref{section:DecodingSummary}. Section
\ref{section:Conclusion} concludes the paper.

\section{Irreducible Polynomial Remainder Codes}
\label{section:CRT} In this section, we quickly recall the Chinese
remainder theorem, the definition of irreducible polynomial
remainder codes, and some basic properties of such codes as in
\cite{YuLoeliger,YuLoeliger2}.

Let $R=F[x]$ be the ring of polynomials over some field $F$. For
any monic polynomial $m(x)\in F[x]$, let $R_m$ denote the ring of
polynomials over $F$ of degree less than $\deg m(x)$ with addition
and multiplication modulo $m(x)$.

\subsection{CRT Theorem and Polynomial Remainder Codes}

\begin{theorem}[Chinese Remainder Theorem]
For some integer \mbox{$n>1$}, let $m_0(x), m_1(x), \ldots,
m_{n-1}(x) \in R$ be relatively prime polynomials, and let $M_n(x)
\eqdef \prod_{i=0}^{n-1}m_i(x)$. The mapping
\begin{IEEEeqnarray}{rCl}
   \psi &:& R_{M_n} \rightarrow R_{m_0} \times \ldots \times R_{m_n}: \nonumber\\
    &&  a(x) \mapsto \psi(a) \eqdef \big( \psi_0(a),\ldots,\psi_{n-1}(a) \big)
        \label{eqn:defPsi}
\end{IEEEeqnarray}
with $\psi_i(a) \eqdef a(x) \bmod m_{i}(x)$ is a ring isomorphism.
The inverse mapping is
\begin{equation}
  \psi^{-1} :
   (c_0,\ldots,c_{n-1}) \mapsto
   \sum_{i=0}^{n-1} c_i(x) \beta_i(x)  \bmod M_n(x) \label{eqn:fixedinversemap}
\end{equation}
with coefficients
\begin{equation}
    \beta_i(x) = \frac{M_n(x)}{m_i(x)} \cdot \left(\frac{M_n(x)}{m_i(x)}\right)_{\!\bmod
    m_i(x)}^{-1} \label{eqn:fixedcoefficients}
\end{equation}
where $\big(b(x)\big)_{\!\bmod m_i(x)}^{-1}$ denotes the inverse
of $b(x)$ in $R_{m_i}$.
\end{theorem}

\begin{definition} \label{def:IPRC_Code}
An \emph{irreducible polynomial remainder code} over $F[x]$ is a
set of the form
\begin{equation} \label{eqn:DefIPRC}
  C \eqdef \{ (c_0,\ldots,c_{n-1}) = \psi(a)~\text{for some}~a(x) \in R_{M_k} \}
\end{equation}
where $n$ and $k$ are integers satisfying $1\leq k \leq n$, where
$m_0(x), m_1(x), \ldots, m_{n-1}(x) \in F[x]$ are different monic
irreducible polynomials, and where $M_k(x) \eqdef
\prod_{i=0}^{k-1}m_i(x)$.
\end{definition}

\subsection{Distance and Error Correction}
\label{section:weight and distance}

Let $C$ be a code as in Definition~\ref{def:IPRC_Code}. Let
$y=c+e$ denote a corrupted codeword that the receiver gets to see,
where $c\in C$ is the transmitted codeword corresponding to some
$a(x)\in R_{M_k}$ by (\ref{eqn:DefIPRC}), and where $e$ is an
error pattern.

For any \mbox{$a(x) \in R_{M_n}$}, the \emph{degree weight} of
$\psi(a) = \big( \psi_0(a),\ldots,\psi_{n-1}(a) \big)$ is
\begin{equation}
\wD(\psi(a)) \eqdef \sum_{i:\psi_i(a)\neq 0}\deg m_i(x).
\end{equation}
For any $a(x),b(x) \in R_{M_n}$, the \emph{degree-weighted
distance} between $\psi(a)$ and $\psi(b)$ is
\begin{equation}
 \dD(\psi(a),\psi(b)) \eqdef \wD(\psi(a)-\psi(b)).
\end{equation}
Let
\begin{equation}
N \eqdef \deg M_n(x) = \sum_{i=0}^{n-1} \deg m_i(x)
\end{equation}
and
\begin{equation}
K \eqdef \deg M_k(x) = \sum_{i=0}^{k-1} \deg m_i(x).
\end{equation}
Then, the degree weight of any nonzero codeword $\psi(a)$
(\mbox{$a(x)\in R_{M_k}$}, \mbox{$a(x)\neq 0$}) satisfies
\begin{equation} \label{eqn:MinDegreeWeight}
\wD(\psi(a)) > N-K
\end{equation}
and the minimum degree-weighted distance of $C$ satisfies
\begin{equation} \label{eqn:MinDegreeDistance}
\dminD(C) > N-K.
\end{equation}
If $C$ also satisfies the \emph{Ordered-Degree Condition}
\begin{equation} \label{eqn:OrderedDegreeCondition}
\deg m_0(x)\leq \deg m_1(x) \leq \ldots \leq \deg m_{n-1}(x),
\end{equation}
then the Hamming weight of any nonzero codeword $\psi(a)$
(\mbox{$a(x)\in R_{M_k}$}, \mbox{$a(x)\neq 0$}) satisfies
$\wH(\psi(a)) \geq n-k+1$ and the minimum Hamming distance of $C$
is $\dminH(C)= n-k+1$.

An error-only decoding algorithm, which is guaranteed to correct
all the error patterns of $\wD(e)<\dminD/2$ and also the error
patterns of $\wH(e)<\dminH/2$ if the code satisfies
(\ref{eqn:OrderedDegreeCondition}), was proposed in
\cite{YuLoeliger,YuLoeliger2} to deal with the error
$e=(e_0,e_1,\ldots,e_{n-1})$ with unknown error positions (i.e.
$e_i\neq 0,0 \leq i\leq n-1,$ are unknown). Moreover, an efficient
interpolation formula was also proposed in
\cite{YuLoeliger,YuLoeliger2} to recover $a(x)$ from $y=c+e$ when
the positions $i$ of $e_i\neq 0$ are all known.

In the following, we consider the problem where only some (rather
than all) positions $i$ of $e_i \neq 0$ are known before decoding.

\section{Error-and-Erasure Decoding}
\label{section:ErrorAndErasure}

In this section, we present three possible approaches to joint
error-and-erasure decoding of the code $C$ as in
Definition~\ref{def:IPRC_Code}.

Let $y=c+e$ denote a corrupted codeword, 
where $c=(c_0,c_1,\ldots, c_{n-1})\in C$ and where
$e=(e_0,e_1,\ldots, e_{n-1})$ is an error pattern. Let $S_{e}
\subset \{0,1,\ldots,n-1\}$ denote the set of positions~$i$ of
$e_i\neq 0$. Let $S_{\rho} \subset S_{e}$ denote the set of known
positions $i$ of $e_i\neq 0$ and let $S_{\tau} \eqdef S_{e}
\setminus S_{\rho}$ denote the set of the unknown positions $i$ of
$e_i\neq 0$.

\subsection{A Modified-Transform Approach}
A first approach, as observed by Shiozaki\cite{Shiozaki}, is to
reduce the joint error-and-erasure decoding of such codes  to the
error-only decoding of the shortened codes.
Specifically, let ${S}\eqdef \{0,1,\ldots,n-1\}\setminus
S_{\rho}$, let $M_{{S}}(x) \eqdef \prod_{i\in {S}} m_i(x)$, and
let $\bigotimes_{i\in {S}} R_{m_i}$ denote the direct product of
the rings $R_{m_i}$ with $i\in {S}$. Moreover, let
$\tilde{c}\eqdef \{c_i\}$ with $i\in {S}$, i.e., $\tilde{c}$ is
the shortened codeword of $c$. It then follows from Theorem~1 that
the mapping ${\phi} : R_{M_{S}} \rightarrow  \bigotimes_{i\in S}
R_{m_i}$ is a ring isomorphism. The inverse mapping is
\begin{equation}
  {\phi}^{-1} :
   \tilde{c} \mapsto
   \sum_{i \in S} c_i(x) \tilde{\beta}_i(x)  \bmod M_{S}(x)
\end{equation}
with interpolating basis
\begin{equation} \label{eqn:modifiedinversemap}
    \tilde{\beta}_i(x) = \frac{M_S(x)}{m_i(x)} \cdot \left(\frac{M_S(x)}{m_i(x)}\right)_{\!\bmod
    m_i(x)}^{-1}
\end{equation}
We can then use the error-only decoding algorithms as in
\cite{YuLoeliger,YuLoeliger2} to decode $\tilde{c}$. This approach
requires, however, a lot of re-computation of
(\ref{eqn:modifiedinversemap}) and thus greatly increases the
decoding complexity, as the case for Reed-Solomon codes
\cite{LinChenTruong}.

In the following two subsections, we propose two other approaches
which avoid the re-computation (\ref{eqn:modifiedinversemap}) and
use the fixed transform $\psi^{-1}$ and the fixed ${\beta}_i(x)$
in (\ref{eqn:fixedinversemap}) and (\ref{eqn:fixedcoefficients}).

\subsection{A Fixed-Transform Approach~I}
\label{ApproachI}

Recall that $y=c+e$. Let $Y(x)=a(x)+E(x)$ denote the pre-image
$\psi^{-1}(y)$ of $y$ with $\psi^{-1}$ as in
(\ref{eqn:fixedinversemap}), where $a(x)=\psi^{-1}(c)$ of $\deg
a(x)<K$ and where $E(x) = \sum_{\ell=0}^{N-1} E_\ell\, x^{\ell}$
denotes the pre-image $\psi^{-1}(e)$ of $e$.

Let
\begin{eqnarray}
\Lambda_e(x) \eqdef \prod_{i \in S_{e}}m_i(x) \label{eqn:el}
\end{eqnarray}
be the unique monic error locator polynomial of the smallest
degree $\deg \Lambda_e(x)= \wD(e)$ \cite{YuLoeliger,YuLoeliger2}.
With
\begin{equation}
\Lambda_\rho(x) \eqdef \prod_{i\in S_{\rho}}m_i(x)
\label{eqn:el.rho}
\end{equation}
and
\begin{equation}
\Lambda_\tau(x) \eqdef \prod_{i\in S_{\tau}}m_i(x),
\label{eqn:el.tau}
\end{equation}
(\ref{eqn:el}) can then be written as
$\Lambda_e(x)=\Lambda_\rho(x) \Lambda_\tau(x), \label{el.rho.tau}$
and the key equation in Theorem~6 of \cite{YuLoeliger} can be
written as
\begin{equation}
      A(x)M_n(x)=\Lambda_\rho(x) \Lambda_\tau(x)E(x).   \label{eqn:KeyEqu.0}
\end{equation}
Now let
\begin{equation}\label{eqn:defHatE}
\hat{E}(x)\eqdef \Lambda_\rho(x)E(x).
\end{equation}
and
\begin{equation}\label{eqn:defHatY}
\hat{Y}(x)\eqdef
\Lambda_\rho(x)Y(x)=\Lambda_\rho(x)a(x)+\hat{E}(x).
\end{equation}
\begin{theorem}\label{theorem:KeyEquationI}
The polynomial (\ref{eqn:el.tau}) satisfies
\begin{equation}
      A(x)M_n(x)=\Lambda_\tau(x)\hat{E}(x)         \label{eqn:KeyEquI.1}
\end{equation}
for some polynomial $A(x)\in F[x]$ of degree smaller than $\deg
\Lambda_e(x)=\Lambda_\rho(x)+\Lambda_\tau(x)$. Conversely, if some
polynomial $G(x)\in F[x]$ satisfies
\begin{equation}
       A(x)M_n(x)=G(x)\hat{E}(x)               \label{eqn:KeyEquI.2}
\end{equation}
for some $A(x)\in F[x]$, then $G(x)$ is a multiple of
$\Lambda_\tau(x)$.
\end{theorem}

\begin{theorem} [Fixed-Transform Interpolation] \label{theorem:ErasureCorrectionI}
If $G(x)$ is a multiple of $\Lambda_\tau(x)$ with
\begin{equation}
   \deg G(x)\leq N-K-\deg \Lambda_\rho(x),
\end{equation}
then
\begin{equation} \label{eqn:LocatorBasedInterpolationI}
a(x)=\frac{\hat{Y}(x)G(x) \bmod M_n(x)}{\Lambda_\rho(x)G(x)}
\end{equation}
\eproofnegspace
\end{theorem}
Theorems \ref{theorem:KeyEquationI} and
\ref{theorem:ErasureCorrectionI} follow easily from Theorems~6
and~7 of \cite{YuLoeliger}. Since $\Lambda_\rho(x)$ is given,
(\ref{eqn:LocatorBasedInterpolationI}) implies that $a(x)$ can be
computed immediately once $\Lambda_\tau(x)$ is known.

\subsection{A Fixed-Transform Approach II}
\label{ApproachII}

Recall that $M_n(x) \eqdef \prod_{i=0}^{n-1}m_i(x)$. Let
\begin{equation}\label{eqn:defTildeMn}
\tilde{M}_n(x)\eqdef M_n(x)/\Lambda_\rho(x),
\end{equation}
which is of degree $\deg\tilde{M}_n(x)=N-\deg\Lambda_\rho(x)$. We
then have the following analog of Theorems
\ref{theorem:KeyEquationI} and \ref{theorem:ErasureCorrectionI}.
\begin{theorem}\label{theorem:KeyEquationII}
The polynomial (\ref{eqn:el.tau}) satisfies
\begin{equation}
      A(x)\tilde{M}_n(x)=\Lambda_\tau(x)E(x)         \label{eqn:KeyEquII.1}
\end{equation}
for some polynomial $A(x)\in F[x]$ of degree smaller than $\deg
\Lambda_e(x)=\Lambda_\rho(x)+\Lambda_\tau(x)$. Conversely, if some
polynomial $G(x)\in F[x]$ satisfies
\begin{equation}
       A(x)\tilde{M}_n(x)=G(x)E(x)               \label{eqn:KeyEquII.2}
\end{equation}
for some $A(x)\in F[x]$, then $G(x)$ is a multiple of
$\Lambda_\tau(x)$.
\end{theorem}

\begin{theorem} [Fixed-Transform Interpolation] \label{theorem:ErasureCorrectionII}
If $G(x)$ is a multiple of $\Lambda_\tau(x)$ with
\begin{equation}
   \deg G(x)\leq N-K-\deg \Lambda_\rho(x),
\end{equation}
then
\begin{equation} \label{eqn:LocatorBasedInterpolationII}
a(x)=\frac{Y(x)G(x) \bmod \tilde{M}_n(x) }{G(x)}.
\end{equation}
\eproofnegspace
\end{theorem}
The two theorems follow also easily from Theorems~6~and~7 of
\cite{YuLoeliger}. From (\ref{eqn:LocatorBasedInterpolationII}),
$a(x)$ can be computed immediately once $\Lambda_\tau(x)$ is
known.

In the following two sections, we will investigate the use of a
modified gcd algorithm to solve the modified key equations
(\ref{eqn:KeyEquI.2}) and (\ref{eqn:KeyEquII.2}).

\section{Solving (\ref{eqn:KeyEquI.2}) by the Extended GCD Algorithm}
\label{section:exgcdI}

It is known that an extended gcd algorithm can be used to solve a
key equation and compute an error locator polynomial
\cite{YuLoeliger,YuLoeliger2}, which is also one of the standard
ways of decoding Reed-Solomon codes \cite{Roth}. We now adapt this
approach to solve the modified key equation (\ref{eqn:KeyEquI.2}).

\subsection{An Extended GCD Algorithm}  \label{section:ExtendedGCDAlgorithmI}

In this subsection, we assume that $\hat{E}(x) \neq 0$ is fully
known; in the next subsection, we state the modifications that are
required when $\hat{E}(x)$ is only partially known. We prefer the
following gcd algorithm \cite{YuLoeliger,YuLoeliger2}.

\begin{trivlist}
\item{} \noindent
{\bf Extended GCD Algorithm I}\\
Input: $M_n(x)$ and $\hat{E}(x)$. \\
Output: polynomials \mbox{$s(x) \text{~and~} t(x) \in F[x]$}, cf.
Theorem \ref{theorem:GCD Output.1}.
\begin{pseudocode}
\npcl[gcdline:Liner]  $r(x) := M_n(x)$\\
\npcl[gcdline:Linetilder]  $\tilde{r}(x) := \hat{E}(x)$ \\
\npcl[gcdline:Lines]    $s(x) := 1$ \\
\npcl[gcdline:Linetx]   $t(x) := 0$ \\
\npcl   $\tilde{s}(x) := 0$ \\
\npcl   $\tilde{t}(x) := 1$ \\
\npcl   \pkw{loop begin} \\
\npcl   \> $i := \deg r(x)$ \\
\npcl[gcdline:Assignj]   \> $j := \deg \tilde{r}(x)$ \\
\npcl[gcdline:BeginWhile]   \> \pkw{while} $i \geq j$ \pkw{begin}\\
\npcl[gcdline:ri]   \> \> $q(x):=\frac{r_i}{\tilde{r}_j}~x^{i-j}$ \\
\npcl[gcdline:Updater]   \> \> $r(x):=r(x)-q(x)\cdot \tilde{r}(x)$ \\
\npcl[gcdline:Updates]   \> \> $s(x):=s(x)-q(x)\cdot \tilde{s}(x)$ \\
\npcl[gcdline:Updatet]  \> \> $t(x):=t(x)-q(x)\cdot \tilde{t}(x)$ \\
\npcl   \> \> $i :=\deg r(x)$ \\
\npcl[gcdline:EndWhile]   \> \pkw{end} \\
\npcl[gcdline:BeginStoppingIf]   \> \pkw{if} $\deg r(x)=0$ \pkw{begin}\\
\npcl[gcdline:Return]   \> \> \pkw{return} $s(x)$, $t(x)$ \\
\npcl[gcdline:BeginStoppingEnd]   \> \pkw{end} \\
\npcl[gcdline:Swapr]   \> $(r(x),\tilde{r}(x)) := (\tilde{r}(x),r(x))$ \\
\npcl[gcdline:Swaps]   \> $(s(x),\tilde{s}(x)) := (\tilde{s}(x),s(x))$ \\
\npcl[gcdline:Swapt]   \> $(t(x),\tilde{t}(x)) := (\tilde{t}(x),t(x))$ \\
\npcl   \pkw{end}
\end{pseudocode}
\vspace{-3ex} \hfill $\Box$
\end{trivlist}
It is easily verified that the standard loop invariant \cite{Roth}
holds also for this gcd algorithm:
\begin{equation}
r(x) = s(x)\cdot M_n(x)+t(x)\cdot \hat{E}(x)
\label{eqn:LoopInvariantrst}
\end{equation}
holds between lines \ref{gcdline:Assignj}
and~\ref{gcdline:BeginWhile} and between lines
\ref{gcdline:EndWhile} and~\ref{gcdline:BeginStoppingIf}.
\begin{theorem}[GCD Output]\label{theorem:GCD Output.1}
When the algorithm terminates, we have both
\begin{equation} \label{eqn:GcdOutputST}
s(x)\cdot M_n(x)+t(x)\cdot \hat{E}(x)=0.
\end{equation}
and
\begin{equation}\label{eqn:RetuenTx}
t(x)=\tilde{\gamma} \Lambda_\tau(x)
\end{equation}
for some scalar $\tilde{\gamma}\in F$.
\end{theorem}

In fact, $\hat{E}(x) \neq 0$ is not fully known, but with the
modification in the following subsection, the Extended GCD
Algorithm~I can be still used to compute $t(x)=\gamma
\Lambda_\tau(x)$.

\subsection{Modifications for Partially Known $E(x)$}
\label{section:ModifiedGCDAlgorithm.1}

Recall that $Y(x)=a(x)+E(x)$. Since $\deg a(x) < K$ the receiver
knows the coefficients $E_{K}, E_{K+1},\ldots, E_{N-1}$ of $E(x)$.
It follows from (\ref{eqn:defHatE}) and (\ref{eqn:defHatY}) that
the upper $N-K$ coefficients of $\hat{E}(x)$, obtained from
$\hat{Y}(x)$, are also known, which can then be used to compute
$t(x)=\gamma \Lambda_\tau(x)$ as follows.
\begin{trivlist}
\item{} \noindent \noindent {\bf Partial GCD Algorithm~I} \\
Input: $M_n(x)$ and $\hat{Y}(x)$. \\
Output: \mbox{$r(x), s(x), t(x)$}, cf.
Theorem~\ref{theorem:PartialGCDAlogI} below.

The algorithm is the same as the Extended GCD Algorithm~I of
Section~\ref{section:ExtendedGCDAlgorithmI} except for the
following changes:

\begin{itemize}

\item Line~\ref{gcdline:Linetilder}: $\tilde{r}(x) := \hat{Y}(x)$

\item Line~\ref{gcdline:BeginStoppingIf}:
  \begin{equation} \label{eqn:ModifiedStopping1}
  \text{\textbf{if}  $\deg r(x) < \deg t(x)+\deg \Lambda_\rho(x)+K$ \textbf{begin}}
  \end{equation}
  or alternatively
  \begin{equation} \label{eqn:ModifiedStopping2}
  \text{\textbf{if}  $\deg r(x) < (N+K+\deg \Lambda_\rho(x))/2$
  \textbf{begin}}
  \end{equation}

\end{itemize}
\vspace{-0ex} \hfill $\Box$
\end{trivlist}
\begin{theorem}\label{theorem:PartialGCDAlogI}
If $\Lambda_\tau(x)$ satisfies
\begin{equation}
 \deg \Lambda_\tau(x)\leq (N-K-\deg\Lambda_\rho)/2,   \label{cond:errorbound}
\end{equation}
then the Partial GCD Algorithm I (with either
(\ref{eqn:ModifiedStopping1}) or (\ref{eqn:ModifiedStopping2}))
returns the same polynomials $s(x)$ and $t(x)$ (after the same
number of iterations) as the Extended GCD Algorithm~I of
Section~\ref{section:ExtendedGCDAlgorithmI}. Moreover, the
returned $r(x)$ is such that
\begin{equation} \label{eqn:PartialGcdOutputI}
r(x)=t(x)\Lambda_\rho(x)a(x).
\end{equation}
\end{theorem}
Note that $a(x)$ can be recovered directly from
(\ref{eqn:PartialGcdOutputI}).

\section{Solving (\ref{eqn:KeyEquII.2}) by the Extended GCD Algorithm}
\label{section:exgcdII}

\subsection{An Extended GCD Algorithm}  \label{section:ExtendedGCDAlgorithmII}

The following Extended GCD Algorithm~II, which is fully described
for clarity and ease of reference, is the same as the Extended GCD
Algorithm~I in Section \ref{section:ExtendedGCDAlgorithmI} except
having different input polynomials. We first assume that $E(x)
\neq 0$ is fully known, and then in the next subsection, we state
the required modifications when $E(x)$ is partially known.

\begin{trivlist}
\item{} \noindent
{\bf Extended GCD Algorithm II}\\
Input: $\tilde{M}_n(x)$ and $E(x)$. \\
Output: polynomials \mbox{$s(x) \text{~and~} t(x) \in F[x]$}.
\begin{pseudocode}
\npcl[gcdline:Liner]  $r(x) := \tilde{M}_n(x)$\\
\npcl[gcdline:Linetilder]  $\tilde{r}(x) := E(x)$ \\
\npcl[gcdline:Lines]    $s(x) := 1$ \\
\npcl[gcdline:Linetx]   $t(x) := 0$ \\
\npcl   $\tilde{s}(x) := 0$ \\
\npcl   $\tilde{t}(x) := 1$ \\
\npcl   \pkw{loop begin} \\
\npcl   \> $i := \deg r(x)$ \\
\npcl[gcdline:Assignj]   \> $j := \deg \tilde{r}(x)$ \\
\npcl[gcdline:BeginWhile]   \> \pkw{while} $i \geq j$ \pkw{begin}\\
\npcl[gcdline:ri]   \> \> $q(x):=\frac{r_i}{\tilde{r}_j}~x^{i-j}$ \\
\npcl[gcdline:Updater]   \> \> $r(x):=r(x)-q(x)\cdot \tilde{r}(x)$ \\
\npcl[gcdline:Updates]   \> \> $s(x):=s(x)-q(x)\cdot \tilde{s}(x)$ \\
\npcl[gcdline:Updatet]  \> \> $t(x):=t(x)-q(x)\cdot \tilde{t}(x)$ \\
\npcl   \> \> $i :=\deg r(x)$ \\
\npcl[gcdline:EndWhile]   \> \pkw{end} \\
\npcl[gcdline:BeginStoppingIf]   \> \pkw{if} $\deg r(x)=0$ \pkw{begin}\\
\npcl[gcdline:Return]   \> \> \pkw{return} $s(x)$, $t(x)$ \\
\npcl[gcdline:BeginStoppingEnd]   \> \pkw{end} \\
\npcl[gcdline:Swapr]   \> $(r(x),\tilde{r}(x)) := (\tilde{r}(x),r(x))$ \\
\npcl[gcdline:Swaps]   \> $(s(x),\tilde{s}(x)) := (\tilde{s}(x),s(x))$ \\
\npcl[gcdline:Swapt]   \> $(t(x),\tilde{t}(x)) := (\tilde{t}(x),t(x))$ \\
\npcl   \pkw{end}
\end{pseudocode}
\vspace{-3ex} \hfill $\Box$
\end{trivlist}
For this algorithm, the loop invariant
\begin{equation}
r(x) = s(x)\cdot \tilde{M}_n(x)+t(x)\cdot E(x),
\label{eqn:LoopInvariantrstII}
\end{equation}
holds between lines \ref{gcdline:Assignj}
and~\ref{gcdline:BeginWhile} and between lines
\ref{gcdline:EndWhile} and~\ref{gcdline:BeginStoppingIf}.
\begin{theorem}[GCD Output]\label{theorem:GCDOutputII}
When the algorithm terminates, we have both
\begin{equation} \label{eqn:GcdOutputIIST}
s(x)\cdot \tilde{M}_n(x)+t(x)\cdot E(x)=0.
\end{equation}
and
\begin{equation}\label{eqn:RetuenIITx}
t(x)=\tilde{\gamma} \Lambda_\tau(x)
\end{equation}
for some scalar $\tilde{\gamma}\in F$.
\end{theorem}

\subsection{Modifications for Partially Known $E(x)$}
Recall that the known coefficients $E_{K}, E_{K+1},\ldots,
E_{N-1}$ of $E(x)$ can be obtained from $Y(x)$.

\label{section:ModifiedGCDAlgorithmIII.1}
\begin{trivlist}
\item{} \noindent \noindent {\bf Partial GCD Algorithm~II} \\
Input: $\tilde{M}_n(x)$ and $Y(x)$. \\
Output: \mbox{$r(x), s(x), t(x)$}, cf.
Theorem~\ref{theorem:PartialGCDAlogII} below.

The algorithm is the same as the Extended GCD Algorithm~II of
Section~\ref{section:ExtendedGCDAlgorithmII} except for the
following changes:

\begin{itemize}

\item Line~\ref{gcdline:Linetilder}: $\tilde{r}(x) := Y(x)$

\item Line~\ref{gcdline:BeginStoppingIf}:
  \begin{equation} \label{eqn:ModifiedStoppingII.1}
  \text{\textbf{if}  $\deg r(x) < \deg t(x)+K$ \textbf{begin}}
  \end{equation}
  or alternatively
  \begin{equation} \label{eqn:ModifiedStoppingII.2}
  \text{\textbf{if}  $\deg r(x) < (N+K-\deg \Lambda_\rho(x))/2$
  \textbf{begin}}
  \end{equation}

\end{itemize}
\vspace{-1ex} \hfill $\Box$
\end{trivlist}
\begin{theorem}\label{theorem:PartialGCDAlogII}
If the condition (\ref{cond:errorbound}) is satisfied, then the
Partial GCD Algorithm~II (with either
(\ref{eqn:ModifiedStoppingII.1}) or
(\ref{eqn:ModifiedStoppingII.2})) returns the same polynomials
$s(x)$ and $t(x)$ (after the same number of iterations) as the
Extended GCD Algorithm II of
Section~\ref{section:ExtendedGCDAlgorithmII}. Moreover, the
returned $r(x)$ is such that
\begin{equation} \label{eqn:PartialGcdOutputII}
r(x)=t(x)a(x).
\end{equation}
\end{theorem}
Note that $a(x)$ can be recovered directly from
(\ref{eqn:PartialGcdOutputII}).

\section{Summary of decoding}
\label{section:DecodingSummary}

Let us summarize the proposed decoding algorithm and add some
details. The receiver sees $y=c+e$ where $c\in C$ is the
transmitted codeword and $e$ is an error pattern. We thus have
$Y(x) = a(x) + E(x)$ where $Y(x)$, $a(x)$, and $E(x)$ are the
images of $y$, $c$, and $e$ under the fixed transform $\psi^{-1}$
and $\deg a(x) < K$.

\subsection{Decoding using Fixed-Transform Approach I}
\label{DecodingbyApproachI}

By Fixed-Transform Approach~I, we first compute $\hat{Y}(x)$ from
(\ref{eqn:defHatY}), and then run the Partial GCD Algorithm~I. If
(\ref{cond:errorbound}) is satisfied, then the algorithm yields
$s(x)$, $t(x)$ and $r(x)$ that satisfy (\ref{eqn:GcdOutputST}),
(\ref{eqn:RetuenTx}) and (\ref{eqn:PartialGcdOutputI}). We can
then recover $a(x)$ by either of the following methods:

\begin{enumerate}

\item From (\ref{eqn:LocatorBasedInterpolationI}), we have

\begin{equation}\label{eqn:ApproachIaxfromtxHatYx}
a(x)=\frac{t(x)\hat{Y}(x) \bmod M_n(x)}{t(x)\Lambda_\rho(x)}
\end{equation}

(If the numerator of (\ref{eqn:ApproachIaxfromtxHatYx}) is not a
multiple of $t(x)\Lambda_\rho(x)$ or if $\deg a(x) \geq K$, then
decoding failed due to some uncorrectable error.)

\item We can compute 
\begin{equation} \label{eqn:ApproachIaxfromrx}
a(x)=\frac{r(x)}{t(x)\Lambda_\rho(x)}
\end{equation}
according to (\ref{eqn:PartialGcdOutputI}).

(If $t(x)\Lambda_\rho(x)$ does not divide $r(x)$ or if $\deg a(x)
\geq K$, we declare a decoding failure.)

\end{enumerate}

When applied to Reed-Solomon codes, Approach I with the recovery
of $a(x)$ by (\ref{eqn:ApproachIaxfromrx}) is identical to the
algorithm proposed in \cite{LinChenTruong}, but recovering $a(x)$
by (\ref{eqn:ApproachIaxfromtxHatYx}) is new.

\subsection{Decoding using Fixed-Transform Approach~II}
\label{DecodingbyApproachII}

By Fixed-Transform Approach~II, we first compute $\tilde{M}_n(x)$
from (\ref{eqn:defTildeMn}), and then run the Partial GCD
Algorithm~II. If (\ref{cond:errorbound}) is satisfied, then the
algorithm yields $s(x)$, $t(x)$ and $r(x)$ that satisfy
(\ref{eqn:GcdOutputIIST}), (\ref{eqn:RetuenIITx}) and
(\ref{eqn:PartialGcdOutputII}). We can then recover $a(x)$ by
either of the following methods:

\begin{enumerate}

\item From (\ref{eqn:LocatorBasedInterpolationII}), we have

\begin{equation}\label{eqn:ApproachIIaxfromtxYx}
a(x)=\frac{t(x)Y(x) \bmod \tilde{M}_n(x)}{t(x)}
\end{equation}

(If the numerator of (\ref{eqn:ApproachIIaxfromtxYx}) is not a
multiple of $t(x)$ or if $\deg a(x) \geq K$, then decoding failed
due to some uncorrectable error.)

\item We can compute 
\begin{equation} \label{eqn:ApproachIIaxfromrx}
a(x)=\frac{r(x)}{t(x)}
\end{equation}
according to (\ref{eqn:PartialGcdOutputII}).

(If $t(x)$ does not divide $r(x)$ or if $\deg a(x) \geq K$, we
declare a decoding failure.)

\end{enumerate}

Note that Approach II appears to be new even when applied to
decoding Reed-Solomon codes. Note also that Approach II (with the
recovery of $a(x)$ either by (\ref{eqn:ApproachIIaxfromtxYx}) or
by (\ref{eqn:ApproachIIaxfromrx})) is of the same form as the
error-only decoding of \cite{YuLoeliger2}.

In comparison with Approach~I, the gcd algorithm of Approach~II
requires less computation since the input polynomials
$\tilde{M}_n(x)$ and $Y(x)$ of the Partial GCD Algorithm~II have
degrees smaller than the inputs $M_n(x)$ and $\hat{Y}(x)$ of the
Partial GCD Algorithm~I.

\section{Conclusion}
\label{section:Conclusion}

We have extended previous work of the error-only decoding of
irreducible polynomial remainder codes to the joint
error-and-erasure decoding of such codes, for which we have
proposed two fixed-transform approaches. As we have shown, for
each approach, the joint error-and-erasure decoding is carried out
by an efficient gcd algorithm, and is fully compatible in
implementation with the error-only decoding. Of particular
interest is the second approach, which appears to be new even when
specialized to Reed-Solomon codes.

\section{Acknowledgement}
The author is deeply grateful to Prof.~H.-A.~Loeliger for his
encouragement and great support of this work.

\newcommand{\COM}{IEEE Trans.\ Communications}
\newcommand{\COMMag}{IEEE Communications Mag.}
\newcommand{\IT}{IEEE Trans.\ Information Theory}
\newcommand{\JSAC}{IEEE J.\ Select.\ Areas in Communications}
\newcommand{\SP}{IEEE Trans.\ Signal Proc.}
\newcommand{\SPMag}{IEEE Signal Proc.\ Mag.}
\newcommand{\ProcIEEE}{Proceedings of the IEEE}


\begin{thebibliography}{99}



\bibitem{Stone}
J.~J.~Stone, ``Multiple-burst error correction with the Chinese
Remainder Theorem,'' \emph{ J.~SIAM}, vol.~11, pp.\ 74--81, Mar.\
1963.

\bibitem{Reed}
I.~S.~Reed and G.\ Solomon, ``Polynominal codes over certain
finite fields,'' \emph{ J.~SIAM}, vol.~8, pp.\ 300--304, Oct.\
1962.

\bibitem{Bossen}
D.~C.~Bossen and S.~S.\ Yau, ``Redundant residue polynomial
codes,'' \emph{Information and Control}, vol.\ 13, pp.\ 597--618,
1968.

\bibitem{Mandelbaum}
D.~Mandelbaum, ``A method of coding for multiple errors,''
\emph{\IT}, vol.\ 14, pp.\ 518--621, May 1968.

\bibitem{Mandelbaum2}
D.~Mandelbaum, ``On efficient burst correcting residue polynomial
codes,'' \emph{Information and Control}, vol.\ 16, pp.\ 319--330,
1970.

\bibitem{Goppa}
V.~D.~Goppa, ``A new class of linear error-correction codes,''
\emph{Probl. Peredach. Inform.}, vol.\ 6, pp.\ 24--30, Sept. 1970.


\bibitem{Shiozaki}
A.~Shiozaki, ``Decoding of redundant residue polynomial codes
using Euclid's algorithm,'' \emph{ \IT}, vol.\ 34, pp.\
1351--1354, Sep.\ 1988.

\bibitem{YuLoeliger}
J.-H.~Yu and H.-A.~Loeliger, ``On irreducible polynomial remainder
codes,'' \emph{IEEE Int. Symp. on Information Theory, Saint
Petersburg, Russia, July 31--Aug. 5, 2011}.

\bibitem{YuLoeliger2}
J.-H.~Yu and H.-A.~Loeliger, ``On polynomial remainder codes,''
\emph{submitted to \IT}. (available on
http://arxiv.org/abs/1201.1812.)

\bibitem{Gao}
S. Gao, ``A new algorithm for decoding Reed-Solomon codes,'' in
\emph{Communications, Information and Network Security}, V.
Bhargava, H. V. Poor, V. Tarokh, and S.Yoon, Eds. Norwell, MA:
Kluwer, 2003, vol. 712, pp. 55-68.


\bibitem{Fedorenko2}
S. V. Fedorenko, ``Correction to `A simple algorithm for decoding
Ree-Solomon codes and its relation to the Welch-Berlekamp
algorithm','' \emph{ \IT}, vol. IT-52, pp. 1278, Mar. 2006.

\bibitem{LinChenTruong}
T.-C.~Lin, P.-D.~Chen, and T.-K.~Truong ``Simplified procedure for
decoding nonsystematic Reed-Solomon codes over GF($2^m$) using
Euclid's algorithm and the fast Fourier transform,'' \emph{\COM},
vol.\ 57, pp.\ 1588--1592, Jun.\ 2009.

\bibitem{Sugiyama}
Y.~Sugiyama, M.~Kasahara, S.~Hirasawa, and T.~Namekawa, ``A method
for solving key equation for decoding Goppa codes,''
\emph{Information and Control}, vol.\ 27, pp.\ 87--99, 1975.

\bibitem{Roth}
R.~M.~Roth, \emph{Introduction to Coding Theory.} New York:
Cambridge University Press, 2006.


\end{thebibliography}
\end{document}